\documentclass[12pt]{article}

\usepackage{geometry}
\usepackage{amsthm}
\usepackage{amsmath}
\usepackage{amssymb}
\usepackage{graphicx}
\usepackage{subfigure}

\newcommand{\bb}[1]{\mathbb{#1}}            % "blackboard-bold" font (style: double-struck)
\newcommand{\bmat}{\begin{bmatrix}}         % \begin{matrix} (style: brackets)
\newcommand{\emat}{\end{bmatrix}}           % \end{matrix} (style: brackets)

\newcommand{\GL}[2]{\mathrm{GL}(#1,\bb{#2})}
\newcommand{\SU}[1]{\mathrm{SU}(#1)}
\newcommand{\SO}[1]{\mathrm{SO}(#1)}
\newcommand{\uone}{\mathrm{U}(1)}

\newcommand{\vek}[1]{\mathbf{#1}}           % indication of 2-vectors (style: bold)

\newcommand{\ole}{\operatorname{\mathcal{O}}}  % Olesen's function

\newcommand{\defeq}{\mathrel{\mathop:}=}    % equal by definition (style: ":=")
\theoremstyle{remark}
\newtheorem*{remark}{Remark on special functions}

\theoremstyle{remark}

\theoremstyle{remark}
\newtheorem*{remark2}{Remark}

\theoremstyle{plain}
\newtheorem*{unnumberedthm}{Theorem}

\theoremstyle{plain}
\newtheorem{theorem}{Theorem}

\theoremstyle{plain}
\newtheorem{lemma}{Lemma}

\theoremstyle{plain}
\newtheorem*{necesslemma2}{Lemma 2 (``$\Rightarrow$'')}

\theoremstyle{definition}

\theoremstyle{definition}
\newtheorem*{nonumdef}{Definition}
\newcommand{\beq}{\begin{equation}}
\newcommand{\eeq}{\end{equation}}

\newcommand\f{{\tilde{f}}}

\newcommand{\GLtwo}{\GL{2}{C}}
\newcommand{\PSUtwo}{\mathrm{PSU(2,\C)}}

\newcommand{\SUtwo}{\SU{2}}
\newcommand{\Utwo}{\mathrm{U}(2)}
\newcommand{\R}{\bb{R}}
\newcommand{\C}{\bb{C}}
\newcommand{\Z}{\bb{Z}}
\newcommand{\Q}{\bb{Q}}
\newcommand{\N}{\bb{N}}
\newcommand{\bH}{\bb{H}}
\newcommand{\ord}{\mathrm{ord}}
\newcommand{\footnoteremember}[2]{\footnote{#2}\newcounter{#1}\setcounter{#1}{\value{footnote}}}
\newcommand{\footnoterecall}[1]{\footnotemark[\value{#1}]}
\begin{document}

\title{Nonrelativistic Chern-Simons Vortices\\on the Torus}
\author{N. Akerblom$^{1,}\footnote{{\tt nikolasa@nikhef.nl}}$, G. Cornelissen$^{2,}\footnote{{\tt g.cornelissen@uu.nl}}$,\\G. Stavenga$^{3,}\footnote{{\tt stavenga@gmail.com}}$, and J.-W. van Holten$^{1,}\footnote{{\tt t32@nikhef.nl}}$
\\
\\
\normalsize{$^1$ Nikhef Theory Group, Amsterdam, The Netherlands}
\\
\normalsize{$^2$ Department of Mathematics, Utrecht University, The Netherlands}
\\
\normalsize{$^3$ Theoretical Physics Department, Fermilab, Batavia, IL, USA}
}

\maketitle

\begin{abstract}
A classification of all periodic self-dual static vortex solutions of the Jackiw-Pi model is given. Physically acceptable solutions of the Liouville equation are related to a class of functions which we term $\Omega$-quasi-elliptic. This class includes, in particular, the elliptic functions and also contains a function previously investigated by Olesen. Some examples of solutions are studied numerically and we point out a peculiar phenomenon of lost vortex charge in the limit where the period lengths tend to infinity, that is, in the planar limit.
\end{abstract}
\vspace{5ex}

\hfill{NIKHEF/2009-030~~~~~~~}

\hfill{FERMILAB-PUB-09-590-T~~~~~~~}

\clearpage
\tableofcontents

\section{Introduction}
In this paper we study periodic, static vortex solutions of the Jackiw-Pi model \cite{Jackiw:1990mb,Jackiw:1990tz}.\footnote{
For reviews see \cite{Jackiw:1991au,Horvathy:2007ps,Horvathy:2008hd}.} This is a $2+1$-dimensional nonrelativistic 
conformal field theory whose field content consists of a complex scalar field $\Psi$ with non-linear-Schr\"odinger type 
action  minimally coupled to a $\uone$ Chern-Simons gauge field $A_\mu$. Let us begin by reviewing the elements of 
this model.

\subsection{The Jackiw-Pi model}\label{review}
We take as our starting point the action \cite{deKok:2008ge}
\beq\label{action}
\begin{array}{lll}
S[\Psi,A_\mu] & = & \displaystyle{
 \int dx^0 \int d^2x\,\Big\{ - \frac{1}{2} \varepsilon_{ij} \left( A_0 \partial_i A_j + A_i \partial_j A_0
 + A_j \partial_0 A_i \right) }\\
 & & \\
 & &  \displaystyle{ \hspace{6em} 
 +\, i \Psi^*D_0 \Psi - \frac{1}{2}\, (\vek{D}\Psi)^*\cdot(\vek{D}\Psi)- \frac{g^2}{2} |\Psi|^4 \Big\}. }
\end{array}
\eeq
Here $\varepsilon_{12} = - \varepsilon_{21} = +1$, whilst $D_\mu=\partial_\mu-ieA_\mu$ is the gauge covariant derivative 
and {\bf bold} type indicates its spatial 2-vector part. We use the generic notation $x^0$ for the time coordinate, and
apply the summation convention for repeated indices. In the following we define
\beq \label{n.1}
\begin{array}{l}
\displaystyle{ \rho = \Psi^* \Psi, \hspace{2em} J_i = - \frac{i}{2} \left( \Psi^* D_i \Psi - \Psi D_i \Psi^* \right), }\\
 \\
B = \partial_1 A_2 - \partial_2 A_1, \hspace{2em} E_i = \partial_0 A_i - \partial_i A_0.
\end{array}
\eeq
The field equations derived from the action (\ref{action}) then read
\beq \label{n.2}
\begin{array}{l}
B = e \rho, \hspace{2em} E_i = e \varepsilon_{ij} J_j, \\
 \\
\displaystyle{ i D_0 \Psi = - \frac{1}{2}\, \vek{D}^2 \Psi + g^2 |\Psi|^2 \Psi.} 
\end{array}
\eeq
The chiral derivatives
\beq \label{n.3}
D_{\pm} = \frac{1}{\sqrt{2}} \left( D_1 \pm i D_2 \right),
\eeq
satisfy the identities
\beq \label{n.4}
\frac{1}{2}\, \vek{D}^2 = D_- D_+ - \frac{e}{2}\, B = D_+ D_- + \frac{e}{2}\, B.
\eeq
Using equations (\ref{n.2}), the Schr\"{o}dinger equation for $\Psi$ can then be written as 
\beq \label{n.5}
i D_0 \Psi = - D_-D_+ \Psi + \left( g^2 + \frac{e^2}{2} \right) |\Psi|^2 \Psi 
 = - D_+ D_- \Psi + \left( g^2 - \frac{ e^2}{2} \right) |\Psi|^2 \Psi.
\eeq
By a similar argument, the hamiltonian takes the form 
\beq \label{n.6}
H = \int d^2 x \left( \frac{1}{2} |\vek{D} \Psi |^2 + \frac{g^2}{2}\, |\Psi|^4 \right) 
 = \int d^2 x \left( |D_{\pm} \Psi|^2 + \frac{1}{2} (g^2 \pm e^2) |\Psi|^4 \right).
\eeq
Hence there are two possibilities for constructing stationary zero-energy solutions: 
\beq \label{n.7}
\begin{array}{ll}
(I) & D_+ \Psi = 0 \hspace{1em} \mbox{and} \hspace{1em} g^2 + e^2 = 0, \\
 & \\
(II) & D_- \Psi = 0  \hspace{1em} \mbox{and} \hspace{1em} g^2 - e^2 = 0. 
\end{array}
\eeq
By stationary we mean that physical observables such as the particle density and current
are time-independent. This is achieved by separating space and time variables as
\beq \label{n.8}
\Psi = e^{i\omega} \sqrt{\rho},
\eeq
with $\rho$ non-negative and time independent: $\partial_0 \rho = 0$. Any time dependence
therefore resides in the gauge-dependent phase $\omega$. 

Substitution of either of the \emph{Ans\"{a}tze} $(I)$ or $(II)$ for $\Psi$ and the coupling constants 
$(e,g)$ simplifies the Schr\"{o}dinger equation (\ref{n.5}) to
\beq \label{n.9}
i D_0 \Psi = (e A_0 - \partial_0 \omega) \Psi = \mp \frac{e^2}{2}\, |\Psi|^2 \Psi \hspace{1em}
 \Rightarrow \hspace{1em} eA_0 = \partial_0 \omega \mp \frac{e^2}{2}\, \rho.
\eeq
In addition, the real and imaginary parts of either condition $D_{\pm} \Psi = 0$ lead to the
real equations
\beq \label{n.10}
e A_i = \partial_i \omega \pm \varepsilon_{ij} \partial_j \ln \sqrt{\rho},
\eeq
and as a result
\beq \label{n.11}
e B = e \varepsilon_{ij} \partial_i A_j = \mp \Delta \ln \sqrt{\rho}, \hspace{2em} 
\Delta = \partial_1^2 + \partial_2^2.
\eeq
It follows directly that $\rho$ satisfies one of the Liouville equations
\beq \label{liouvilleeq}
\begin{array}{ll}
(I) & D_+ \Psi = 0 \hspace{1em} \Rightarrow \hspace{1em} \Delta \ln \sqrt{\rho} + e^2 \rho = 0, \\
 \\
(II)& D_- \Psi = 0 \hspace{1em} \Rightarrow \hspace{1em} \Delta \ln \sqrt{\rho} - e^2 \rho = 0.
\end{array}
\eeq
The solutions of these equations are respectively of the form \cite{Liouville}
\beq \label{n.13}
\begin{array}{ll}
(I) & \displaystyle{ \rho_f = \frac{4}{e^2}\, \frac{|f^{\prime}|^2}{(1+ |f|^2)^2},}\\
 \\
(II)& \displaystyle{\rho_f = \frac{4}{e^2}\, \frac{|f^{\prime}|^2}{(1- |f|^2)^2},}\end{array}
\eeq
  where $f(z)$ is an analytic function of the complex coordinate
\beq \label{n.14}
z = x + i y,
\eeq
and for physical reasons we make the hypothesis that $f$ have at most isolated singularities (which then automatically are poles; see the discussion in Section \ref{classi}).

Furthermore, boundedness of $\rho$ requires 
$|f|^2 < 1$ for case $(II)$. This immediately implies that there are no relevant non-trivial solutions in 
case $(II)$.

However, case $(I)$ leads to a rich spectrum of vortex-type solutions, depending on the boundary conditions. 
For instance, if we take two-dimensional space to be the plane $\R^2\leftrightarrow\C$, it is necessary to 
require that at infinity $\rho$ tends to zero sufficiently fast. The problem of writing down all static vortex 
solutions in this planar case was solved in a beautiful paper by Horvathy and Yera \cite{Horvathy:1998pe}.

For physical applications, e.g.\ in condensed matter systems, it is also of interest to study static vortex 
solutions in a finite volume with periodic boundary conditions; in  that case one requires
\beq
\rho(z+\omega_i)=\rho(z)
\eeq
for given $\R$-linearly independent complex numbers $\omega_1$, $\omega_2$. This corresponds to studying 
the Jackiw-Pi model in the case where space is a two-dimensional flat torus $\C/(\Z\omega_1+\Z\omega_2)$. 
Apparently, the first one to investigate this situation was Olesen, who gave a remarkable solution on a square 
torus \cite{Olesen:1991dg}.

If one thinks about how periodic boundary conditions are customarily employed in physics, the step from the 
plane to the torus seems innocuous. However, here this is not at all so. The change in topology, in fact, has 
a rather dramatic effect on the allowed rationalized vortex charge\footnote{For a proof of equations 
\eqref{chargeeq} see \cite{Jackiw:1990tz}.}
\beq\label{chargeeq}
q\defeq \frac{|\text{vortex charge}|}{2\pi\times e}=\frac{\text{(total magnetic flux)}\times e}{2\pi}
 = \frac{e^2}{2\pi}\int \rho\, d^2x.
\eeq
First of all, it can be shown that $q$, irrespective of whether we are working on the plane or the torus, must be 
a non-negative integer (see Appendix \ref{quantflux}). But, while from the classification of Horvathy and Yera it 
is immediate that on the plane the charge is always even:
\beq
q=2n,\quad n\in\N_0\quad\text{(vortex charge on the plane)},
\eeq
this is not the case for the Jackiw-Pi model on the torus. Indeed, Olesen's solution is an example of a static 
vortex on the torus with charge
\beq
q=1\quad\text{(Olesen's solution on the torus)}.
\eeq
Moreover, the correspondence between the solutions on the torus and those on the plane is somewhat involved. 
Olesen's solution, for instance, vanishes in the limit where the period lengths tend to infinity, and in Section 
\ref{examples} we give an example of a solution for which the charge $q$ is \emph{halved} as we pass from 
the torus to the plane! The adagium that the limit of a periodic solution, as the periods tend to infinity, gives a planar solution, fails dramatically in the case of Olesen's solution.

\subsection{Classification of vortex solutions} \label{classi}
In complex coordinates (\ref{n.14}) the non-linear wave equation \eqref{liouvilleeq} for positive chirality
fields of type $(I)$ reads
\beq \label{complexliouville}
\bar{\partial} \partial \ln \rho + e^2 \rho = 0,
\eeq
where $\partial\defeq(\partial_1-i\partial_2)/\sqrt{2}$ and $\bar\partial\defeq(\partial_1+i\partial_2)/\sqrt{2}$.

The general solution (\ref{n.13}, $I$) of this equation was discovered a long time ago by Liouville \cite{Liouville}, 
who was led to the study of equations (\ref{n.13}, $I$\&$II$) in connection with his researches on 
the theory of surfaces with constant intrinsic curvature\footnoteremember{metric}{On a surface of constant 
curvature the conformal factor of the metric in isothermal coordinates satisfies the Liouville equation; that is, 
if the metric is $ds^2=\rho\,(dx^2+dy^2)$ with $\rho>0$, then $\rho$ satisfies equation \eqref{liouvilleeq} 
and $e^2$ is equal to the Gaussian curvature $K$ of the surface. In this situation, the case $K<0$ is, of 
course, not excluded and corresponds to solution (\ref{n.13}, $II$). It is known that equation (\ref{liouvilleeq}, $I$) has no nowhere vanishing solution on the torus \cite{Kazdan}. Thus, by necessity, all our torus solutions given below have zeros.} (see also \cite{Bateman,Taoblog}; in \cite{Arthurs} solutions with vanishing boundary 
conditions on a rectangle were investigated). We shall frequently call $\rho_f$ defined by eq.\ (\ref{n.13}, $I$) 
``the density associated with $f$.''

For our purposes, on physical grounds, we make the \emph{hypothesis} that $f$ is to have at most isolated 
singularities. This is because we want to interpret $\rho_f$ as a soliton (a vortex) and this interpretation is 
upset when $f$ has a non-isolated singularity.\footnote{Indeed, we may conjecture that if $f$ has a non-isolated singularity, then its associated density $\rho_f$ is unbounded.}

We also demand that $\rho$ be bounded. In fact, we impose the stronger condition that the total particle number
in the spatial domain
\beq\label{intdensity}
\int \rho\, d^2x,
\eeq
proportional to the total magnetic flux carried by vortices, is finite.\footnote{In the plane case the integral extends 
over $\R^2$, whereas in the periodic case it is taken over some elementary cell; say, the closure of the fundamental 
region: $\{t_1\omega_1+t_2\omega_2 \,|\, 0\leq t_1,t_2\leq1\}$.}
\emph{In the case of a periodic $\rho$, boundedness automatically follows from continuity}, as we can interpret 
$\rho$ as living on a compact space (the torus), and in this case, boundedness is all we need for the integral 
\eqref{intdensity} to make sense. For vortices on the plane, one obviously needs to supplant this with a suitable 
decay condition at infinity, see \cite{Horvathy:1998pe}. 

It can be shown:
\begin{lemma}[Horvathy-Yera \cite{Horvathy:1998pe}]\label{hylemma}
Let $\rho_f$ be the density associated with a complex function $f$ having at most isolated singularities. 
If $\rho_f$ is bounded, then the only possible singularities of $f$ are poles, i.e. $f$ is meromorphic in the plane.
\end{lemma}
In the plane case this extends to infinity, so that $f$ is a meromorphic function on the sphere, that is, a 
\emph{rational function}:
\begin{unnumberedthm}[Horvathy-Yera \cite{Horvathy:1998pe}]
Let the density $\rho_f$ associated with $f$ be a vortex solution of the Liouville equation on the plane. Then $f$ is a rational function, i.e. there are polynomials $P(z)$ and $Q(z)$, such that
\[
f(z)=\frac{P(z)}{Q(z)}.
\]
Moreover, the converse is also true.
\end{unnumberedthm}

In the case of the torus, Lemma \ref{hylemma} still holds (since it is a local statement), but boundedness of $\rho_f$ is automatic, and there is no corresponding statement about the behavior of $f$ ``at infinity.''

We now state the analogous classification in the case where $\rho$ is periodic, or, as one could also say, lives 
on a torus:
\begin{theorem}\label{mainresult}
Let $\rho$ be a smooth periodic solution of the Liouville equation \eqref{liouvilleeq} with periods $\omega_1$ and 
$\omega_2$. It follows that $\rho=\rho_f$ for some complex function $f$ (Liouville, \cite{Liouville}) meromorphic 
in the plane (Lemma \ref{hylemma}) which falls into one of the following two cases:
\begin{description}
\item[Case A] There are complex numbers $\mu_1$, $\mu_2$ with $|\mu_i|=1$, such that
\beq
f(z+\omega_i)=\mu_i\, f(z),
\eeq
that is, $f$ is an elliptic function of the second kind with multipliers $\mu_i$ of unit modulus. For the reader's convenience, we repeat the results of \cite{Lu} for such functions in Appendix \ref{secondelliptic}.
\item[Case B] There are complex parameters $z_1,\ldots,z_n$ in the fundamental region of the lattice  $\Z\omega_1+\Z\omega_2$, and complex constants $a_0,\ldots,a_n$, such that
\beq
f(z)=-\frac{\varphi(z)-1}{\varphi(z)+1}\,\ole(z),
\eeq
where
\beq
\varphi(z)=\left[a_0+\sum_{k=1}^{n} a_k\frac{d^k\zeta}{dz^k}(z-z_0)\right]
\frac{\sigma(z-z_0)^n}{\prod_{k=1}^{n}\sigma(z-z_k)}\, e^{\zeta(\omega_1/2)\,z},
\eeq
with $z_0=\frac{\omega_1}{2n}+\frac{1}{n}\sum_{k=1}^n z_k$,
and
\beq
\ole(z)=\frac{\wp_{2\omega_1,2\omega_2}(z)+b}{c\,\wp_{2\omega_1,2\omega_2}(z)+d},
\eeq
for a suitable choice of parameters $b$, $c$, $d$, given in equations \eqref{bd} and \eqref{c}. 
\end{description}
Moreover, the converse is also true: If $f$ falls into one of the two cases above, its associated density $\rho_f$ 
is a periodic solution of the Liouville equation.
\end{theorem}
\noindent
This result is derived in the next section.

\begin{remark}
Our conventions for the appearing special functions are as follows:
\begin{itemize}
\item $\wp_{\omega_1,\omega_2}$ indicates the Weierstrass p-function associated with the lattice $\Omega=\Z\omega_1+\Z\omega_2$.

\item $\zeta=\zeta_{\omega_1,\omega_2}$ and $\sigma=\sigma_{\omega_1,\omega_2}$ are the Weierstrass zeta- 
and sigma-functions.
\end{itemize}
The properties of these functions are given in many textbooks; see, for example, \cite{Ahlfors}. A word of caution: 
In the older literature, e.g. in the standard reference \cite{Whittaker}, $\wp=\wp_{\omega_1,\omega_2}$ often 
denotes the Weierstrass p-function with \emph{half}-periods $\omega_1$, $\omega_2$ (and similarly for 
$\zeta$ and $\sigma$).
\end{remark}

\section{Periodic vortices}\label{periodicvortices}
We now proceed to classify all periodic vortices on a given flat torus (Theorem \ref{mainresult}). To this end, let a lattice $\Omega\subset\C$ be given and suppose it is spanned by $\omega_1, \omega_2$, that is,
\beq
\Omega=\Z\omega_1+\Z\omega_2.
\eeq
As follows from our earlier discussion in Section \ref{review}, the task is to find all smooth solutions $\rho$ of the Liouville equation \eqref{liouvilleeq} such that
\beq\label{omegaperiodic}
\rho(z+\omega)=\rho(z)\quad\text{for all }\omega\in\Omega.
\eeq
Suppose we are given such a $\rho$. Then, from \cite{Liouville} and Lemma \ref{hylemma} we know that there is a complex function $f$, meromorphic on the plane, such that
\beq
\rho=\rho_f=\frac{4}{e^2}\,\frac{|f'|^2}{(1+|f|^2)^2},
\eeq
where the prime ${}'$ denotes the derivative with respect to the complex variable $z$.

Let $\omega\in\Omega$ be arbitrary and define the function
\beq
g(z)\defeq f(z+\omega).
\eeq
From equation \eqref{omegaperiodic} and the fact that $g'(z)=f'(z+\omega)$, it follows that
\beq
\rho_f(z)=\rho_g(z)\quad\text{for all }z\in\C.
\eeq
In Appendix \ref{proof} we prove:
\begin{lemma}\label{psutransflemma}\footnote{Another proof has been given by de Kok \cite{deKokprivate}.}
Let $f_1$ and $f_2$ be non-constant meromorphic functions on the plane and suppose that their associated densities $\rho_{f_1}$ and $\rho_{f_2}$ are equal: $\rho_{f_1}=\rho_{f_2}$.

Then there exists a matrix
\begin{equation*}
\gamma=
\bmat
a & b\\
c & d
\emat\in\SUtwo,
\end{equation*}
such that
\beq
f_1(z)=\gamma\cdot f_2(z)\defeq
\bmat
a & b\\
c & d
\emat\cdot f_2(z)
\defeq\frac{a f_2(z) + b}{c f_2(z) + d}.
\eeq
Also, the converse is true \cite{deKok:2008ge,deKok:2008}, even under the weaker hypothesis that
$
\bmat
a & b\\
c & d
\emat\in\Utwo$; that is, if $f_1=V\cdot f_2$ for some $V\in\Utwo$, then $\rho_{f_1}=\rho_{f_2}$.\footnote{For $M=\bmat \alpha & \beta \\ \gamma & \delta\emat\in\GLtwo$ and any complex function $f$ we define $M\cdot f(z)\defeq\frac{\alpha f(z)+\beta}{\gamma f(z)+\delta}$.}
\end{lemma}

Now, from Lemma \ref{psutransflemma} it follows that for any $\omega\in\Omega$, there is a matrix $\gamma_\omega\in\SUtwo$, such that
\beq\label{periodcond}
f(z+\omega)=g(z)=\gamma_\omega\cdot f(z).
\eeq
This matrix is \emph{not} unique in $\SUtwo$, but it is unique in $\PSUtwo$.
We shall call a meromorphic function on the plane $\Omega$-quasi-elliptic if it satisfies condition \eqref{periodcond}.

A trivial {\bf corollary} to Lemma \ref{psutransflemma} is that $\rho=\rho_f$ is periodic with respect to the lattice $\Omega=\Z\omega_1+\Z\omega_2$ if and only if there exists matrices $\delta_{\omega_i}\in\Utwo$, such that $f(z+\omega_i)=\delta_{\omega_i}\cdot f(z)$ for $i=1,2$.

With every matrix
$\gamma=
\bmat
a & b\\
c & d
\emat
\in\SUtwo$ there is naturally associated a certain transformation $T(\gamma)\in\PSUtwo$ from the Riemann sphere $\widehat{\C}$ to itself, namely
\beq
T(\gamma):\, \widehat{\C}\rightarrow\widehat{\C},\quad z\mapsto\frac{az+b}{cz+d}.
\eeq
Since, obviously,
\begin{equation}\label{commutgamma}
T(\gamma_{\omega+\tilde{\omega}})=T(\gamma_{\omega})T(\gamma_{\tilde{\omega}})=T(\gamma_{\tilde{\omega}})T(\gamma_{\omega})\quad\text{for all }\omega, \tilde{\omega}\in\Omega,
\end{equation}
equation \eqref{periodcond} tells us that any $\Omega$-quasi-elliptic function effects a group homomorphism
\beq
T:\,\Omega\rightarrow G,\quad \omega\mapsto T(\gamma_\omega),
\eeq
from the lattice $\Omega$ to some abelian subgroup $G$ of $\PSUtwo$. We recall that $\PSUtwo=\SO3$, the group of orientation preserving isometries of the sphere.\footnote{For completeness we mention the 
elementary rule $T(\gamma_1\gamma_2)=T(\gamma_1)T(\gamma_2)$ for all $\gamma_1, \gamma_2\in\Utwo$.} 

Because $\Omega=\Z\omega_1+\Z\omega_2$ is a free module with generators $\omega_1$ and $\omega_2$, $G$ 
is an abelian group with at most two generators $T(\gamma_{\omega_1})$ and $T(\gamma_{\omega_2})$.

Now, the important thing is that the converse part of Lemma \ref{psutransflemma} guarantees that any $\Omega$-quasi-elliptic function will also yield a periodic vortex solution of the Liouville equation. Therefore, the problem of finding all periodic vortex solutions is equivalent to writing down all $\Omega$-quasi-elliptic functions and this is directly related to classifying all abelian subgroups of $\PSUtwo$ with two generators.

There are various ways to classify such subgroups. We will work in $\PSUtwo$ directly, and lift the two generators of the group from $\PSUtwo$ to $\SUtwo$. One can also use the isomorphism with $\SO3$, or consider rotations as quaternions. We shall comment on this later.

By an earlier remark (immediately below equation \eqref{periodcond}), we have the implication
\beq
T(\gamma)=T(\tilde{\gamma})\Rightarrow\gamma=\pm\tilde{\gamma}\quad\text{for all }\gamma,\tilde{\gamma}\in\SUtwo.
\eeq
Then, since the generators $T(\gamma_{\omega_1})$ and $T(\gamma_{\omega_2})$ of $G$ commute,
\begin{equation*}
T(\gamma_{\omega_1})T(\gamma_{\omega_2})=T(\gamma_{\omega_2})T(\gamma_{\omega_1}),
\end{equation*}
it is easy to see that $\gamma_{\omega_1}$ and $\gamma_{\omega_2}$ either commute or anticommute. We will refer to these cases as Case A and Case B, respectively and treat them in turn in the following two sections.
\subsection{Case A: The matrices $\gamma_{\omega_1}$ and $\gamma_{\omega_2}$ commute}
Since $\gamma_{\omega_1}$ and $\gamma_{\omega_2}$ commute, they can simultaneously be put into diagonal form. More precisely, there exists a matrix $U\in\SUtwo$, such that
\beq
\gamma_{\omega_i}=U^\dagger
\bmat
\sqrt{\mu_i} & 0\\
0 & 1/\sqrt{\mu_i}
\emat
U\quad(i=1,2),
\eeq
where the $\mu_i$ are complex numbers of unit modulus: $|\mu_i|=1$.

Let $f$ be $\Omega$-quasi-elliptic and define the function
\begin{equation}
g(z)=U\cdot f(z).
\end{equation}
It follows that
\beq
g(z+\omega_i)=\mu_i\,g(z)\quad(i=1,2),
\eeq
i.e. the function $g$ is a so-called elliptic function of the second kind. There exists a complete classification of all such 
functions (cf. Appendix \ref{secondelliptic}).
Thus, $f$ will be of the form
\beq
f=U^\dagger\cdot g
\eeq
with $g$ some elliptic function of the second kind, and, by Lemma \ref{psutransflemma}, the densities associated with these functions are the same:
\beq
\rho_f=\rho_g.
\eeq

Conversely, if $g$ is a quasi-elliptic function of the second kind with multipliers $\mu_i$ satisfying $|\mu_i|=1$, then its associated density $\rho_g$ is periodic. Indeed, for any such function $g$ there are matrices
\beq
\gamma_{\omega_i}=
\bmat
\sqrt{\mu_i} & 0\\
0 & 1/\sqrt{\mu_i}
\emat\in\SUtwo\quad(i=1,2),
\eeq
with
\beq
g(z+\omega_i)=\gamma_{\omega_i}\cdot g(z)\quad(i=1,2),
\eeq
and the claim immediately follows from the corollary to Lemma \ref{psutransflemma}.

\subsection{Case B: The matrices $\gamma_{\omega_1}$ and $\gamma_{\omega_2}$ anticommute}
If our matrices $\gamma_{\omega_1}$ and $\gamma_{\omega_2}$ anticommute, we can diagonalize one of them and anti-diagonalize the other. Specifically, there is a matrix $U\in\SUtwo$, such that
\beq
\gamma_{\omega_1}=U^\dagger
\bmat
-i & 0\\
0 & i
\emat
U,\quad
\gamma_{\omega_2}=U^\dagger
\bmat
0 & -\lambda\\
\lambda^{-1} & 0
\emat
U,
\eeq
for some complex $\lambda$ with $|\lambda|=1$.
Now put
\beq
M\defeq
\bmat
1 & 0\\
0 & i\,\lambda
\emat,
\eeq
whence
\beq
\gamma_{\omega_1}=U^\dagger M^\dagger
\bmat
-i & 0\\
0 & i
\emat MU,\quad
\gamma_{\omega_2}=U^{\dagger}M^{\dagger}
\bmat
0 & i\\
i & 0
\emat MU,
\eeq
which is to say
\beq
\gamma_{\omega_1}=V^\dagger
\bmat
-i & 0\\
0 & i
\emat V,\quad
\gamma_{\omega_2}=V^\dagger
\bmat
0 & i\\
i & 0
\emat V
\eeq
for some $V=MU\in\Utwo$. 

Let us briefly digress to remark on the subgroup $G$ of $\PSUtwo$ generated by $T(\gamma_{\omega_1})$ and $T(\gamma_{\omega_2})$.

If we define
\beq
a\defeq T(\gamma_{\omega_1}):\,
z\mapsto
\bmat
-i & 0\\
0 & i
\emat\cdot z\,,\quad\quad
b\defeq T(\gamma_{\omega_2}):\,
z\mapsto
\bmat
0 & i\\
i & 0
\emat\cdot z,
\eeq
and
\beq
c\defeq a\circ b:\,z\mapsto
\bmat
0 & 1\\
-1 & 0
\emat\cdot z,
\eeq
we get the composition table of the famous \emph{Vierergruppe} $V=\Z_2\times\Z_2$:
\beq
\begin{array}{c|cccc}
\circ & 1 & a & b & c\\
\hline
    1 & 1 & a & b & c\\
    a & a & 1 & c & b\\
    b & b & c & 1 & a\\
    c & c & b & a & 1
\end{array},
\eeq
where $1$ denotes the identity transformation $z\mapsto z$. Our subgroup $G$ is isomorphic to $\Z_2\times\Z_2$!

Coming back to our classification problem, it follows that any $\Omega$-quasi-elliptic function $f$ is of the form
\beq
f=V^\dagger\cdot g,
\eeq
where $g$ is a function meromorphic in the plane satisfying
\beq\label{oleperiodic}
g(z+\omega_1)=-g(z),\quad g(z+\omega_2)=1/g(z).
\eeq
Conversely, from the corollary to Lemma \ref{psutransflemma} it is plain that the density $\rho_f$ associated with any such $f$ is periodic, for there are matrices $M_1,M_2\in\Utwo$, such that $f(z+\omega_i)=M_i\cdot f(z)$ for $i=1,2$. Moreover, $\rho_f=\rho_g$.

We now proceed to classify all meromorphic functions in the plane which satisfy the period condition \eqref{oleperiodic}.
Suppose $g_0(z)$ is some such function satisfying equation \eqref{oleperiodic} and let $g(z)$ be any other such function. Put
\beq
f(z)\defeq g(z)/g_0(z).
\eeq
Then
\beq
f(z+\omega_1)=f(z),\quad f(z+\omega_2)=1/f(z).
\eeq
If we define
\beq
\varphi(z) \defeq U^\dagger\cdot f(z)
\eeq
with
\beq
U\defeq
\bmat
-1 & 1 \\
1 & 1
\emat,
\eeq
it follows that
\beq
\varphi(z+\omega_1)=\varphi(z),\quad \varphi(z+\omega_2)=-\varphi(z);
\eeq
therefore, $\varphi(z)$ is some multiplicative quasi-elliptic function with $\mu_1=1$, $\mu_2=-1$.

From Appendix \ref{secondelliptic}, we find that there are complex constants
\beq
a_0,\dots,a_n\in\C,
\eeq
and parameters
\beq
z_1,\dots,z_n\in\{t_1\omega_1+t_2\omega_2\,|\,0\leq t_1,t_2<1\}
\eeq
in the fundamental domain of the lattice $\Omega=\Z\omega_1+\Z\omega_2$, such that
\beq
\varphi(z)=\left[a_0+\sum_{k=1}^{n} a_k\frac{d^k\zeta}{dz^k}(z-z_0)\right]
\frac{\sigma(z-z_0)^n}{\prod_{k=1}^{n}\sigma(z-z_k)}\, e^{\zeta(\omega_1/2)\,z},
\eeq
where
\beq
z_0=\frac{\omega_1}{2n}+\frac{1}{n}\sum_{k=1}^n z_k.
\eeq
Therefore, $g$ is of the form
\beq
g(z)=\big[U\cdot \varphi(z)\big]\,g_0(z)=
-\frac{\varphi(z)-1}{\varphi(z)+1}\,g_0(z).
\eeq
Conversely, any such function $g$ satisfies the conditions \eqref{oleperiodic}.

It remains to give some $g_0$ satisfying equation \eqref{oleperiodic}. Inspired by Olesen's special
solution \cite{Olesen:1991dg}, we make the \emph{Ansatz} 
\beq\label{oleansatz}
g_0(z)=\ole(z)\defeq\frac{\wp_{2\omega_1,2\omega_2}(z)+b}{c\,\wp_{2\omega_1,2\omega_2}(z)+d}.
\eeq 
We have the general formulas \cite{Olesen:1991df}
\beq
\wp(z+\omega_1)=e_1+\frac{(e_1-e_2)(e_1-e_3)}{\wp(z)-e_1},\quad
\wp(z+\omega_2)=e_2+\frac{(e_2-e_1)(e_2-e_3)}{\wp(z)-e_2},
\eeq
with $\wp\equiv\wp_{2\omega_1,2\omega_2}$, $e_1\defeq\wp(\omega_1)$, $e_2\defeq\wp(\omega_2)$, 
and $e_3\defeq-(e_1+e_2)$.
Using these formulas and demanding that $g_0$ satisfy \eqref{oleperiodic}, we can choose the parameters 
$b$, $c$, and $d$ in our \emph{Ansatz} \eqref{oleansatz} appropriately.
With the help of a computer algebra system (Mathematica) we have found that
\beq\label{bd}
b=\frac{-e_2^2+c^2(-2e_1+e_2)}
{1+c^2},\quad
d=\frac{c(-2e_1+e_2-c^2e_2)}
{1+c^2},
\eeq
with
\beq\label{c}
c=\sqrt{
\frac{-3e_1+2\sqrt{(e_1-e_2)(2e_1+e_2)}}
{e_1+2e_2}
}
\eeq
will do, as long as $e_1+2e_2\neq0$.\footnote{It turns out to be immaterial which branches we choose for the square roots. In this sense, the choice of parameters is essentially unique.} Indeed, $e_1+2e_2=0$ only in the limit where our torus degenerates into a cylinder and this is excluded.
This concludes our proof of Theorem \ref{mainresult}.

\subsection{The abstract underlying group}

We now explain how to refine our classification from a different
perspective, using the isomorphism $\PSUtwo \cong \SO3 \cong \bH^1$
with the  different model groups of space rotations, and unit
quaternions $\bH^1$.   Let us denote by $G$ the subgroup (in any of
these models) generated by $\gamma_{\omega_1}$ and $\gamma_{\omega_2}$.
Then $G$ is an abelian group of rotations, which is intrinsically
attached to the vortex solutions of the torus Jackiw-Pi model. We call
the abstract isomorphism type of this group the \emph{type of the
vortex solution.}

We denote by $\Q$ the set of rational numbers. As usual, we call a real number
irrational if it is not rational. We call two real numbers linear
dependent over $\Q$ (abbreviated ``LD'') if one is a rational multiple
of the other (and linear independent otherwise).

Suppose our rotations are around the same axis, one through an angle $2
\pi \theta$, the other through an angle $2 \pi \theta'$. If one of
$\theta$ and $\theta'$, say $\theta$, is rational with denominator $m$, then its associated rotation generates a
cyclic subgroup $\Z_m$ of $G$ of order $m$. If then $\theta'$ is
irrational, we find that $G \cong \Z_m \times \Z$ (where it is
possible that $m=1$, in which case $G$ is infinite cyclic: $G \cong \Z$). If both $\theta$ and $\theta'$
are rational with denominators $m$ and $n$, say, then $G$ is a cyclic
group of order the least common multiple $\textrm{lcm}(m,n)$ of $m$
and $n$, that is, $G \cong \Z_{\textrm{lcm}(m,n)}$, a finite cyclic
group (possibly \emph{trivial}, which corresponds to genuinely elliptic
functions). Finally, if $\theta$ and $\theta'$ are both irrational and linearly independent over $\Q$, the corresponding rotations generate a
group $G \cong \Z \times \Z$, but if they are linearly dependent over
$\Q$, they generate a group $G \cong \Z$.

Suppose now that $G$ consists of two commuting rotations around
different axes. It is easy to show (e.g., using the unit quaternion
picture, in which a rotation around an axis $\vec{v}=(v_1,v_2,v_3)$
through an angle $2\theta$ is represented by $\cos \theta + \sin
\theta(v_1 i + v_2 j + v_3 k)$) that the only pair of
commuting rotations are two rotations of $180^\circ$ around two
orthogonal axes, and then, abstractly, the group $G$ is the
\emph{Vierergruppe}. Also, up to an isometry of space, we can assume
that the axes are in a fixed position, so this group $G$ can be
conjugated in $\SO3$ into standard form.

Thus, we see that Case A corresponds to rotations around the same
axis,  whereas Case B corresponds to the \emph{Vierergruppe} of two
rotations around two different axes.

We have summarized the preceding discussion in Table \ref{tablerot}. In this table, we
denote by $\ord(\mu)$ the multiplicative order of a complex number
$\mu$ in $\C^*$, i.e., the smallest positive integer $N$ for which $\mu^N=1$
(and we put $\ord(\mu)=\infty$ if no such integer exists). We call two
complex numbers $\mu_1$ and $\mu_2$ \emph{multiplicatively dependent}
(abreviated ``MD'') if there exist integers $N_1$ and $N_2$ such that
$\mu_1^{N_1} = \mu_2^{N_2}$. We denote a space rotation around an axis
$\vec{v}$ through an angle $\theta$ by $R_{\vec{v}}(\theta)$. Note again
that in this table $m$ and $n$ are integers, so
$\Z_{\mathrm{lcm}(m,n)}$ can be the trivial group (if $m=n=1$), and
$\Z_m \times \Z$ can be an infinite cyclic group $\cong\Z$ (if $m=1$).

\begin{table}
\label{tablerot}
\begin{tabular}{|lll|}
\hline
in $\SUtwo$ & in $\SO3$ & type   \\
\hline
\textbf{Case A: commuting} & \textbf{Same rotation axes} & \\
$\bullet\ \ord(\mu_1)=m$ and $\ord(\mu_2)=n$ & $\langle R_{\vec{v}}(2
\pi/m), R_{\vec{v}}(2 \pi/n) \rangle$ &  $\Z_{\mathrm{lcm}(m,n)} $ \\
$\bullet\ \ord(\mu_1)=m$ and $\ord(\mu_2)=\infty$ & $\langle
R_{\vec{v}}(2 \pi/m), R_{\vec{v}}(2 \pi \theta) \rangle, \ \theta
\notin \Q$  & $\Z_m \times \Z$  \\
$\bullet\ \ord(\mu_1)=\ord(\mu_2)=\infty$ MD & $\langle R_{\vec{v}}(2
\pi \theta), R_{\vec{v}}(2 \pi \theta') \rangle, \ \theta, \theta'
\notin \Q$ LD & $\Z$   \\
$\bullet\ \ord(\mu_1)=\ord(\mu_2)=\infty$ not MD & $\langle
R_{\vec{v}}(2 \pi \theta), R_{\vec{v}}(2 \pi \theta') \rangle, \
\theta, \theta' \notin \Q$ not LD & $\Z \times \Z$   \\
\hline
\textbf{Case B: anticommuting} & \textbf{Orthogonal rotation axes} & \\
& $\langle R_{\vec{v}}(\pi), R_{\vec{w}}(\pi) \rangle \ (\vec{v} \perp
\vec{w})$ & $\Z_2 \times \Z_2$   \\
\hline
\end{tabular}
\caption{Possible ``types'' of vortex solutions on the torus ($m,n$ are integers).}
\end{table}

\section{Examples}\label{examples}
\subsection{Flux loss and flux conservation for elliptic function solutions}

A brief glance at Theorem \ref{mainresult} will convince the reader that, in particular, the densities associated with elliptic functions furnish examples of periodic vortices (take $\mu_1,\mu_2=1$ in Case A). The \emph{type} of these solutions is trivial. 

A function $f$ is elliptic with respect to the lattice $\Omega=\Z\omega_1+\Z\omega_2$ precisely if it can be expressed as
\beq
f(z)=R_1(\wp(z))+\wp'(z)\,R_2(\wp(z)),
\eeq
where $R_1$, $R_2$ are rational functions and $\wp\equiv\wp_{\omega_1,\omega_2}$.

It is easy to see that if we put $\omega_i\rightarrow t\, \omega_i$ ($i=1,2$) and take the limit $t\rightarrow +\infty$, then (compare \cite{Siegel:1969}, pp. 85 ff.)
\beq \label{rat}
f(z)\rightarrow R_1(z^{-2})-2z^{-3}\,R_2(z^{-2}).
\eeq
That is, in the limit where we remove the periodic boundary conditions (the planar limit), $f$ tends to a rational function.
Since any rational function can be written in the form (\ref{rat}) for some rational function $R_2$, any rational function can arise in this way as the limit of an elliptic function. Thus, in this way we obtain all static vortices on the plane.

\begin{figure}[ht]
\centering
\includegraphics[width=12cm]{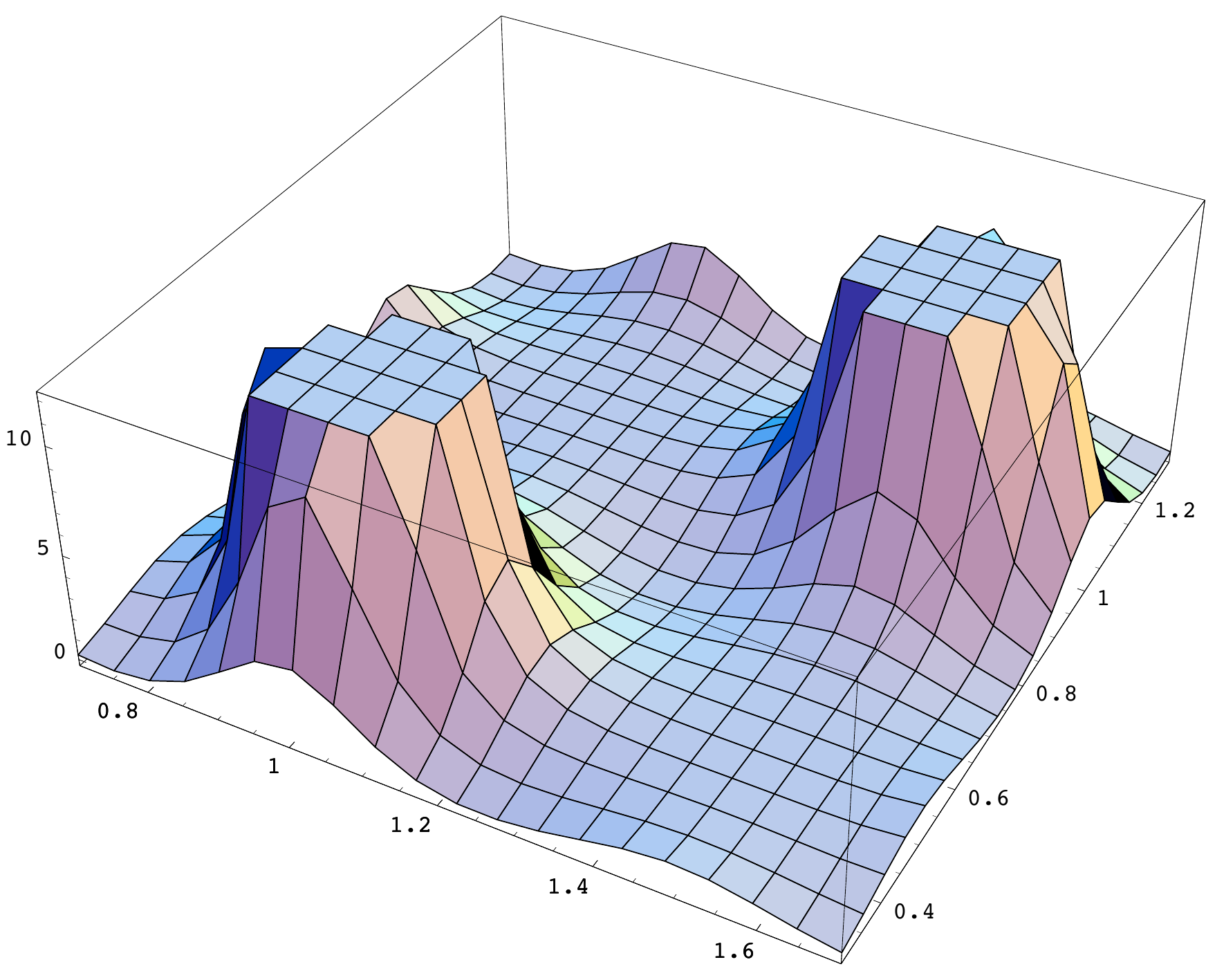}
\caption{Plot of the density (in units of $1/e^2$) associated with the function ${\wp_{t,it}'(z)}/{\wp_{t,it}(z)}$ for $t=1$ 
in the cell $0.7 \leq x \leq 1.7$, $0.3 \leq y \leq 1.3$. Large values of the density have been clipped.}
\label{oddelliptic}
\end{figure}

\paragraph{An elliptic solution with flux loss.} 
Let $\rho_{f_t}$ ($t>0$) be the density associated with the function
\beq
f_t(z)\defeq\frac{\wp_{t,it}'(z)}{\wp_{t,it}(z)}.
\eeq
(We are dealing with the torus $\C/(\Z t+\Z it)$.) Figure \ref{oddelliptic} shows a plot of this density for $t=1$.
Numerical integration suggests that for the rationalized charge $q_{\mathrm{torus}}$ associated with this solution 
($t>0$ finite) we have
\beq
q_{\mathrm{torus}}=\frac{e^2}{2\pi}\int_F \rho_{f_t}\,d^2x=4\quad(F\defeq[0,t]\times[0,t]).
\eeq
Now, the planar limit of $f_t$ is
\beq
f_t(z)\rightarrow \frac{-2}{z}\quad\text{for }t\rightarrow +\infty,
\eeq
and it is well known that the charge associated with this is
\beq
q_{\mathrm{plane}}=\frac{e^2}{2\pi}\int_{\R^2}\rho_{z\mapsto-2/z}\,d^2x=2=\frac{1}{2}\,q_{\mathrm{torus}}.
\eeq
We therefore have the surprising result that, \emph{in passing from the torus to the plane, some charge of a vortex 
can get lost.}

\paragraph{An elliptic solution with no flux loss.} That this need not always happen is shown by the example of the density associated with $g(z)=\wp_{t,it}(z)$. Here, 
the charge in the planar limit is the same as on the torus, namely $=4$.

\subsection{Relatives of Olesen's solution}
\begin{figure}[ht]
\centering
\includegraphics[width=12cm]{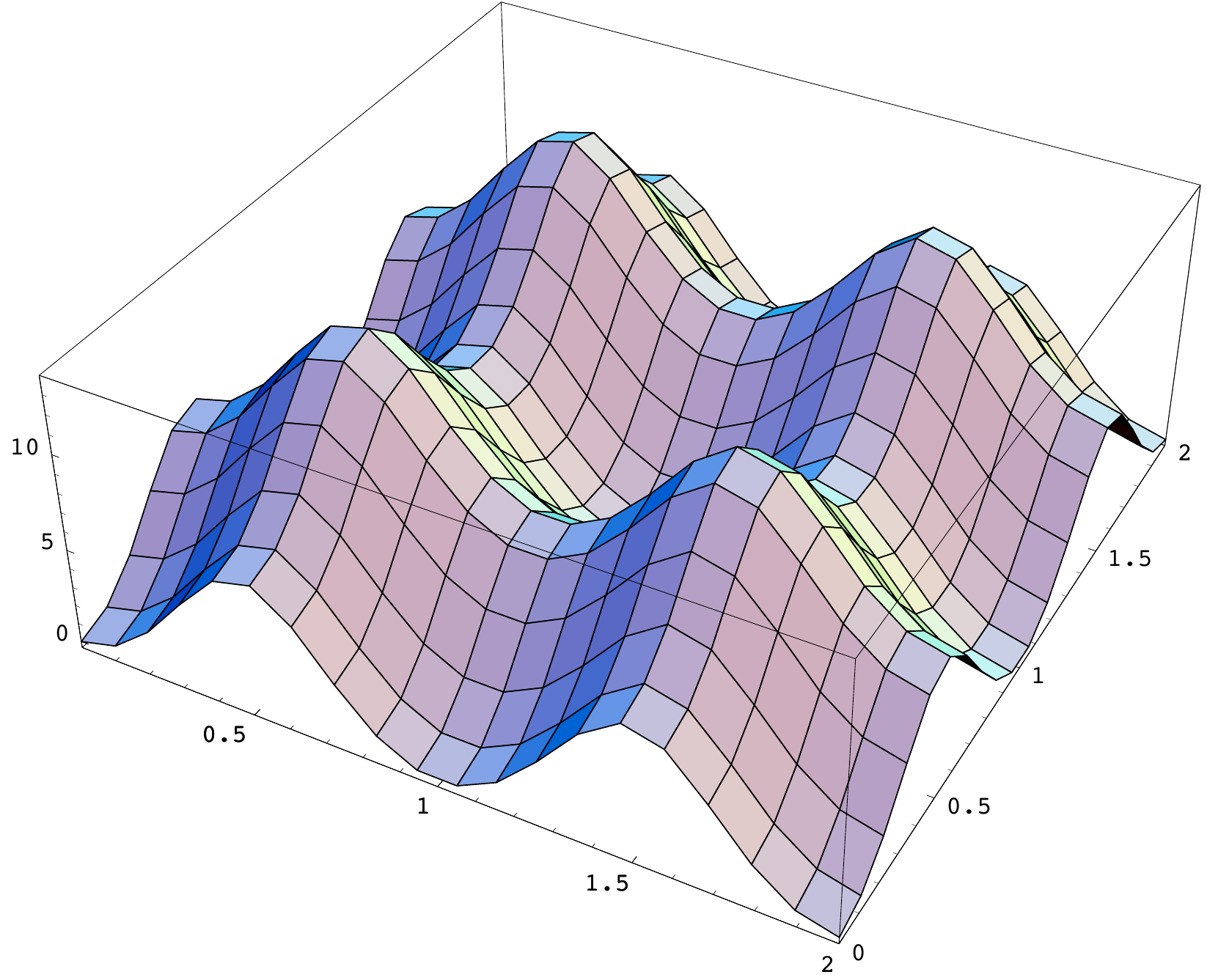}
\caption{Plot of Olesen's density (in units of $1/e^2$) on a $2\times2$ grid of cells where the fundamental domain 
is $[0,1)\times[0,1)$.}\label{olefig}
\end{figure}

In \cite{Olesen:1991dg} Olesen investigated a periodic vortex with charge $q=1$. In our language, this solution is 
associated with the function $\ole{(z)}$ (equation \eqref{oleansatz}) for the square lattice $\Omega=\Z t+\Z i t$ 
with $t>0$. In Figure \ref{olefig} we have plotted this density for $t=1$.

\begin{figure}[ht]
\centering
\subfigure[$t=.5$]{
\includegraphics[scale=.4]{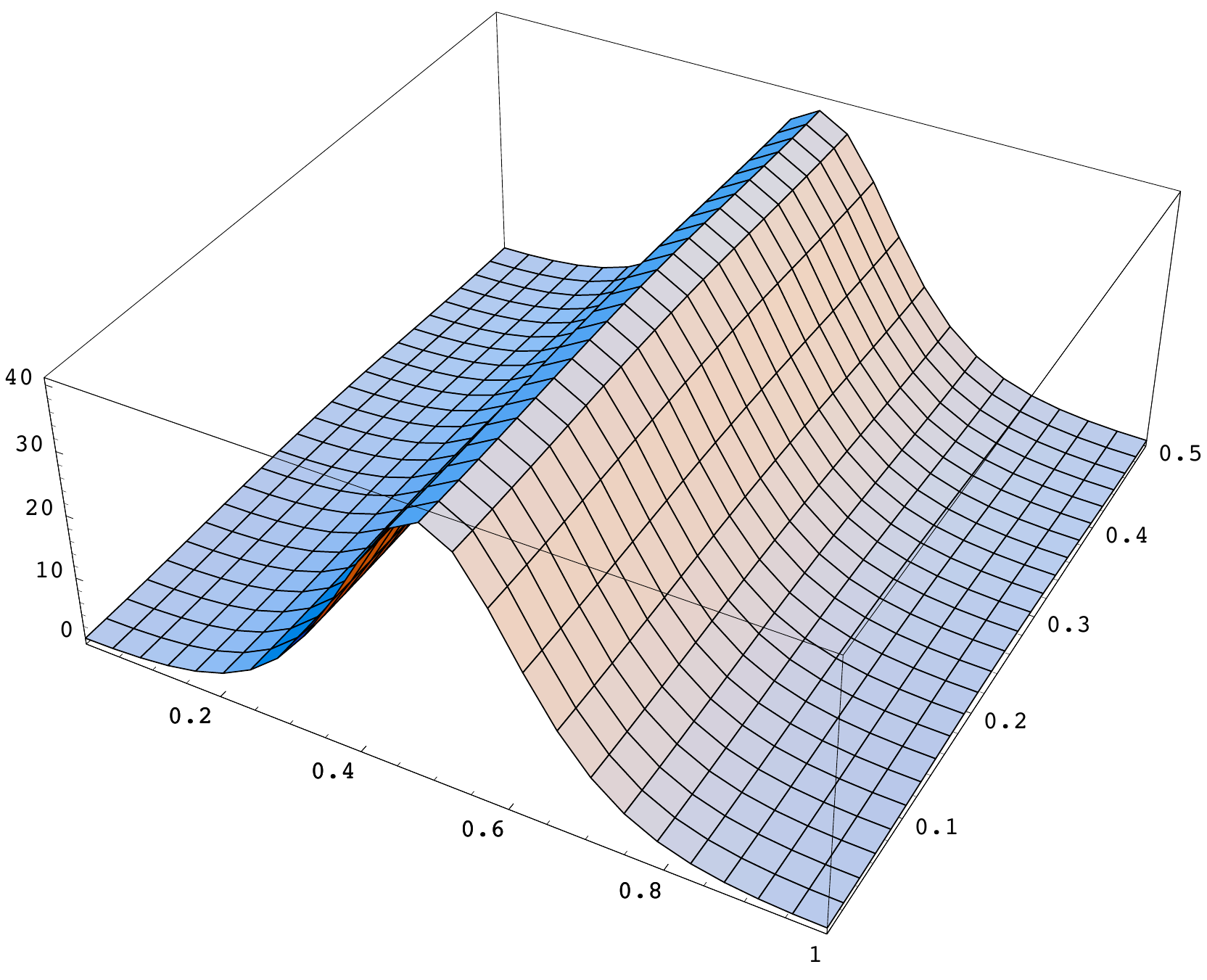}
\label{fig:subfig1}
}
\subfigure[$t=.75$]{
\includegraphics[scale=.4]{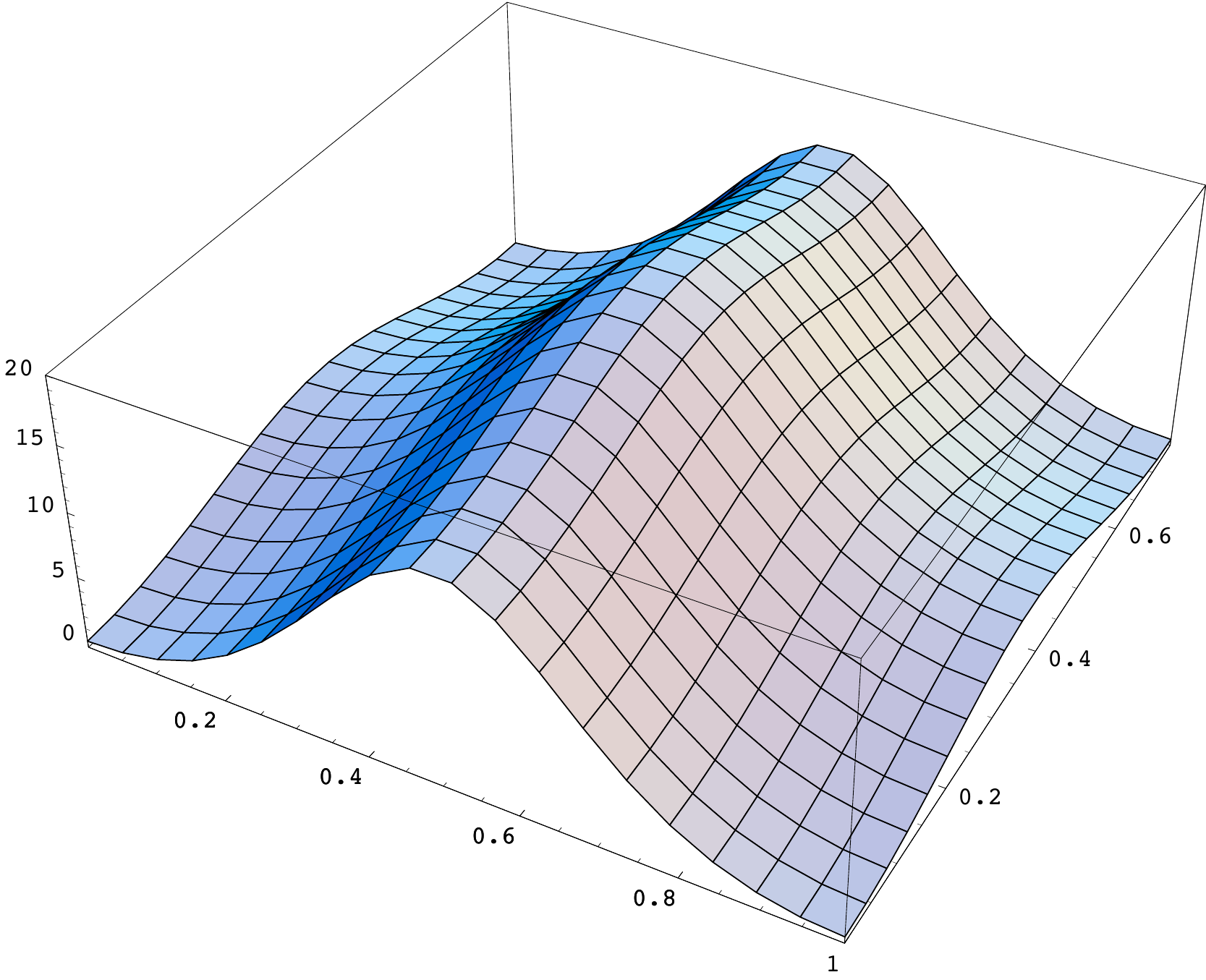}
\label{fig:subfig2}
}
\subfigure[$t=1$]{
\includegraphics[scale=.4]{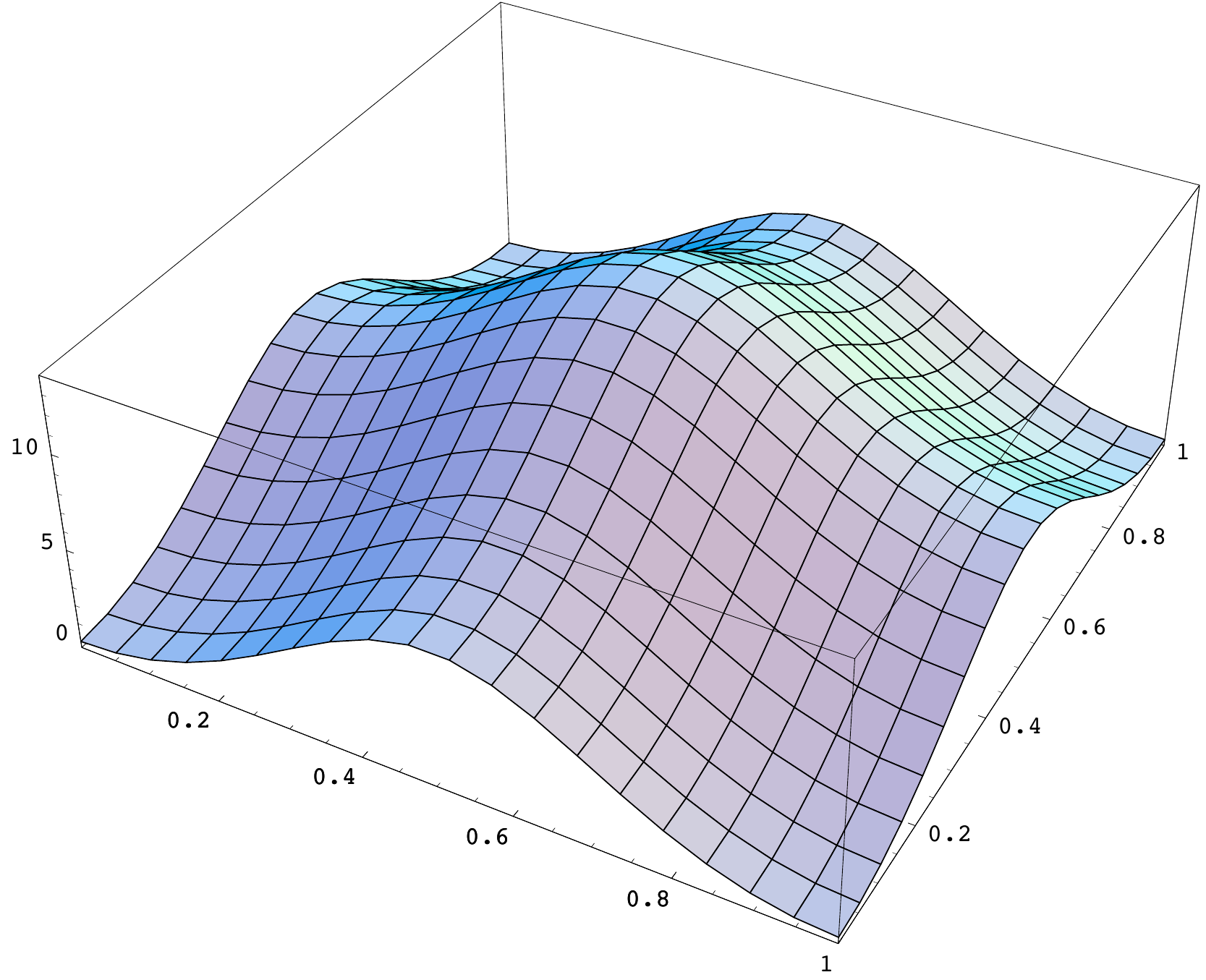}
\label{fig:subfig3}
}
\label{fig:subfigureExample}
\caption{Density (in units of $1/e^2$) associated with the function $\ole(z)$ on a sequence of rectangular tori with 
lattice $\Omega=\Z+\Z it$ for: \subref{fig:subfig1} $t=.5$, \subref{fig:subfig2} $t=.75$, and \subref{fig:subfig3} 
$t=1$.}\label{sequence}
\end{figure}

We can also look at the density associated with $\ole{(z)}$ on arbitrary tori $\C/(\Z\omega_1+\Z\omega_2)$. For instance, Figure \ref{sequence} shows the density for a sequence of lattices $\Omega=\Z+\Z i t$, where successively $t=.5, .75, 1$. Note how the drempel-like structure\footnote{``Drempel'' is a Dutch word which, amongst other things, denotes a speed bump.} deforms to the lump of Figure \ref{olefig} as the rectangle approaches a square. From numerical integration we know that all these vortices have charge $q=1$ and the same appears to be true for tori where the fundamental region is a true parallelogram.

What is the planar limit of the density associated with $\ole{(z)}$? It is easy to see that for a square fundamental region, $\ole{(z)}$ approaches a constant as the period lengths tend to positive infinity, that is, in this limit, the associated density approaches $0$. It appears likely that the same is also true for more general fundamental regions.

\section{Summary and discussion}

In this paper we have studied the Jackiw-Pi model with periodic boundary conditions, which amounts to
solving the Liouville equation on the torus. Physically, these solutions describe a two-dimensional periodic 
lattice of charged vortices with quantized magnetic flux. As first discussed in \cite{Jackiw:1990mb,Jackiw:1990tz} the existence of vortex solutions requires a delicate tuning of the coupling parameters: 
the electric charge and the strength of the self-interaction. Surprisingly, it seems that this tuning is not 
destroyed by quantum fluctuations \cite{Bergman:1994,deKok:2008ge,deKok:2008}; on the 
contrary, the tuning is precisely the condition for which the $\beta$-functions of the model vanish and there 
is no scale-dependence of the parameters, at least at one-loop order. 

On the torus, the spectrum of fluxes of the vortices differs from the planar case; it is richer in that it allows
both odd and even integer fluxes. This is possible because periodic functions on the plane do not vanish
at infinity, as required for the solutions on the infinite plane. However, it also implies that the limit of the
torus to the infinite plane is singular and can change the flux associated with a certain solution. We have
presented explicit examples of this phenomenon. This observation may be relevant also in other field
theories with soliton solutions, e.g.\ the Skyrme model as an effective theory for the bound states in QCD. 

It is amusing to note that our physical classification of vortices on the torus has a purely mathematical consequence 
having to do with the geometrical content of the Liouville equation:
We can interpret our density $\rho$ as the conformal factor of a metric
on a punctured torus, with punctures exactly at the zeros of $\rho$. Our
classification theorem then gives all sufficiently smooth
metrics of constant Gaussian curvature $K=e^2>0$ on punctured
tori in explicit form.\footnote{On an unpunctured torus, there are no such metrics,
compare footnote \footnoterecall{metric}.}
From our physical arguments in Appendix \ref{quantflux} it also follows that the properly normalized 
integral \eqref{chargeeq} of the conformal factor over the torus is always a non-negative integer.

In reference \cite{Horvathy:1999qj} a topological interpretation of the charge of vortex solutions on the plane was given. It would clearly be interesting to obtain an analogous interpretation for the theory on the torus and we believe that the remarks in Appendix \ref{quantflux} could constitute the first steps in that direction.

\section*{Acknowledgments}
We are indebted to P. Horvathy for correspondence and comments, and to C. Hill, S. Moster, E. Plauschinn and B. Schellekens for 
helpful discussions. Two of us (N.\ Akerblom and J.-W. van Holten) have their work supported by the Dutch Foundation 
for Fundamental Research on Matter (FOM). NA also thanks the Max-Planck-Institute for Physics (Munich) for hospitality during the final stage of this paper.
Fermilab is operated by Fermi Research Alliance, LLC under Contract No. DE-AC02-07CH11359 with the US Department of Energy.

\begin{appendix}

\section{Proof of Lemma \ref{psutransflemma}}\label{proof}
Here we prove Lemma \ref{psutransflemma} of Section \ref{periodicvortices}. We only need to supply the proof of the ``$\Rightarrow$''-direction; for the ``$\Leftarrow$''-direction see \cite{deKok:2008ge,deKok:2008}. For clarity, let us repeat the statement (in slightly altered notation):
\begin{necesslemma2}
Let $f$ and $\f$ be non-constant meromorphic functions on the plane and suppose that their associated densities $\rho_{f}$ and $\rho_{\f}$ are equal: $\rho_{f}=\rho_{\f}$,
where
\beq
\rho_f(z)=\frac{4}{e^2}\,\frac{|f'(z)|^2}{(1+|f(z)|^2)^2},
\eeq
and analogously for $\rho_\f$.

Then there exists a matrix
\begin{equation*}
\gamma=
\bmat
a & b\\
c & d
\emat\in\SUtwo,
\end{equation*}
such that
\beq
\f(z)=\gamma\cdot f(z)\defeq
\bmat
a & b\\
c & d
\emat\cdot f(z)
\defeq\frac{a f(z) + b}{c f(z) + d}.
\eeq
\end{necesslemma2}

\begin{proof}
Stereographic projection  $\pi: \, S^2 \rightarrow \widehat{\C}_w$ gives a bijection between the sphere $S^2$ and the extended complex $w$-plane $\widehat{\C}_w$. In this way, the round metric on the sphere,  $ds^2_{S^2}$, induces a distance function $d_U$ on $\widehat{\C}_w$, for which the distance between two points $w_1, w_2\in\widehat{\C}_w$ is given by
\beq
d_U(w_1,w_2)=\inf_\Gamma \int_0^1 \frac{|\Gamma'(t)|}{1+|\Gamma(t)|^2}\,dt,
\eeq
where the infimum is over all curves $\Gamma : \, [0,1] \rightarrow \widehat{\C}_w$ with $\Gamma(0)=w_1$, $\Gamma(1)=w_2$. The orientation preserving isometry group of the sphere, $\SO{3}$, is mapped by $\pi$ to the orientation preserving isometries of $\widehat{\C}_w$ equipped with the distance $d_U$, which is $\PSUtwo$. 

For a meromorphic function on the plane $\C_z$, define a quasi-distance $d_f$ by
\beq\label{defdis}
d_f(z_1,z_2)\defeq d_U(f(z_1),f(z_2)).
\eeq 
(We call this a quasi-distance since, although it is positive and satisfies the triangle inequality, it is degenerate in the sense that points $z_1$, $z_2$ at distance zero are not necessarily equal, but rather satisfy $f(z_1)=f(z_2)$.)

The hypothesis of the theorem concerning equality of densities implies that for every $z_1,z_2 \in \C_z$, we have
\beq\label{dis}
d_f(z_1,z_2)=d_\f(z_1,z_2).
\eeq

We now define a map $\iota : \, \widehat{\C}_w \rightarrow \widehat{\C}_w$ by 
\beq
\iota(w)\defeq\f(f^{-1}(w)).
\eeq
First of all, this is well-defined. Indeed, if $f(z_1)=f(z_2)=:w$, then the definition \eqref{defdis} implies that $d_f(z_1,z_2)=0$. Further, equation \eqref{dis} implies that $d_\f(z_1,z_2)=0$, and, again by definition \eqref{defdis}, we obtain $\f(z_1)=\f(z_2)$.
Our claim is that $\iota(w)$ is an isometry of $\widehat{\C}_w$ equipped with $d_U$.

It is surjective, since $f$ and $\f$ are not constant. Indeed, for any two points $w_1, w_2\in\C_w$, we have 
\beq
d_U(\iota(w_1),\iota(w_2))=d_\f(f^{-1}(w_1),f^{-1}(w_2))=d_f(f^{-1}(w_1),f^{-1}(w_2))=d_U(w_1,w_2).
\eeq

Also, $\iota$ is orientation-preserving since $f$ is meromorphic.

Hence, $\f=T(f)$ for some orientation preserving isometry $T$ of $\widehat{\C}_w$, that is $T\in\PSUtwo$, whence there is a matrix
\beq
\gamma=
\bmat
a & b\\
c & d
\emat\in\SUtwo,
\eeq
such that
\beq
\f=\gamma\cdot f.
\eeq
\end{proof}
\begin{remark2}
It is clear that this lemma can be used for determining the precise structure of the moduli space of self-dual static 
vortices of the Jackiw-Pi model on the plane. For, according to Horvathy and Yera \cite{Horvathy:1998pe}, any 
such vortex with flux $\Phi=4\pi N/e$ is given by a density $\rho_f$, where $f$ is a rational function
\beq
f(z)=\frac{P(z)}{Q(z)},\quad \deg{P}<\deg{Q}=N.
\eeq
Therefore, every such solution has $4N$ moduli but, obviously, they are not all independent. Rather, by our result, 
the moduli space is some kind of quotient
\begin{equation*}
\C^{2N}/\PSUtwo.
\end{equation*}
The invariant theory of $\PSUtwo$ is well-studied, see e.g.\ \cite{Springer}. 
We leave the problem of working out the physical implications in detail for the future \cite{inprep}.
\end{remark2}

\section{Quantization of flux}\label{quantflux}

We comment here on the quantization of flux of static vortex solutions of the Jackiw-Pi model.

For the theory on the plane, this quantization is best seen a posteriori from the results of Horvathy and Yera \cite{Horvathy:1998pe}.
For the time being, an analogous result on the torus is, however, not available \cite{inprep}. That is, given a solution from the classification Theorem \ref{mainresult} we cannot say at the moment, without resorting to numerical integration, what its associated flux is.

Therefore, we now proceed to give a more general argument supporting the claim that the flux is also quantized in the torus case.

The boundary conditions of the Jackiw-Pi model on a spacetime of the form $\R\times T^2$, where $T^2=\C/\Omega$ for some lattice $\Omega$, are somewhat subtle.
Naively, one would write the gauge potential $A$ as a 1-form on the torus, 
which would lead to
\begin{equation}
\int_{T^2} B=\int_{T^2} dA=\int_{\partial T^2=\emptyset} A=0,
\end{equation}
in contradiction to the solutions with a non-vanishing magnetic flux. The resolution to this puzzle is
of course analogous to the Dirac monopole, where we need multiple gauge patches
to describe the solution; in other words, $A$ in reality is a section of a bundle.

However, because we are dealing with a torus, we can also pull back the gauge connection to
the plane, where the gauge potential can be written as a 1-form. The boundary
conditions are then implemented by periodicity of the fields $\rho$, $E$, and $B$,
which translates to the equations
\begin{equation}
\begin{split}
\Psi(x+\omega_i)=e^{i\theta_i(x)}\Psi(x),\\
A(x+\omega_i)=A(x)+d\theta_i(x),\\
\end{split}
\end{equation}
where $\Omega=\Z\omega_{1}+\Z\omega_2$, that is, our lattice is generated by $\omega_1$ and $\omega_2$.

Now we can use gauge transformations in the plane to set the phase $\theta_2$
to zero, and then we are left with a single phase $\theta_1$. It is easy to
show that under translation by $\omega_2$ we have
\begin{equation}
e^{i\theta_1(x)}=e^{i\theta_1(x+\omega_2)},
\end{equation}
and thus $\theta_1(x+\omega_2)=\theta_1(x)+2\pi n$. This means that
the total magnetic flux through the torus is
\begin{equation}
\int_F B=\int_F dA=\int_{\partial F} A,
\end{equation}
where
\beq
F\defeq\{t_1\omega_1+t_2\omega_2\,|\,0\leq t_1,t_2<1\}
\eeq
is the fundamental domain of the torus in the plane. The boundary
integral is the integral along the parallelogram where the two sides in
the direction of $\omega_1$ cancel, due to periodicity of $A$ in $\omega_2$.
However, the sides in the direction of $\omega_2$ do not cancel, due to the
non-periodicity caused by $\theta_1$. The difference between the two sides
is given by
\begin{equation}
\int_{\partial F} A=\int_0^{\omega_2} d\theta_1=2\pi n.
\end{equation}
Therefore, the total magnetic flux is quantized in units of $2\pi$. The topology
of the principal $\uone$ gauge bundle over the torus is that of a twisted
3-torus with twist $n$.

\section{Elliptic functions of the second kind}\label{secondelliptic}
For easy reference we repeat here the results of \cite{Lu}, p.\ 154 concerning elliptic functions of the second kind (=multiplicative quasi-elliptic functions) specialized to the needs of the present paper (see also \cite{Forsyth:1893, EDM}).
\begin{nonumdef}
Let $\Omega=\Z\omega_1+\Z\omega_2$ be a lattice. A function $f$ which is meromorphic in the plane is said to be an \emph{elliptic function of the second kind with multipliers of unit modulus}, if there exist complex numbers $\mu_1,\mu_2$, with $|\mu_1|,|\mu_2|=1$, such that
\beq
f(z+\omega_i)=\mu_i\,f(z)\quad(i=1,2).
\eeq
\end{nonumdef}

\begin{unnumberedthm}[Lu \cite{Lu}]
A function $f$ which is meromorphic in the plane is an elliptic function of the second kind with multipliers $\mu_1,\mu_2$ of unit modulus if and only if there are complex constants
\beq
a_0,\dots,a_n\in\C,
\eeq
and parameters
\beq
z_1,\dots,z_n\in\{t_1\omega_1+t_2\omega_2\,|\,0\leq t_1,t_2<1\},
\eeq
such that
\beq
f(z)=\left[a_0+\sum_{k=1}^{n} a_k\frac{d^k\zeta}{dz^k}(z-z_0)\right]
\frac{\sigma(z-z_0)^n}{\prod_{k=1}^{n}\sigma(z-z_k)}\, e^{\lambda\,z},
\eeq
where
\beq
\lambda=\frac{1}{\pi i}\,(\gamma_2\,\eta_1-\gamma_1\,\eta_2),
\eeq
and
\beq
z_0=\frac{1}{2n\pi i}\,(\gamma_2\,\omega_1-\gamma_1\,\omega_2)+\frac{1}{n}\sum_{k=1}^n z_k.
\eeq
Here, $\eta_i\defeq\zeta_{\omega_1,\omega_2}(\omega_i/2)$ and $\gamma_i\defeq\log{\mu_i}$ ($i=1,2$). (The branch of $\log{\mu_i}$ can be chosen arbitrarily.)
\end{unnumberedthm}

\end{appendix}

\bibliographystyle{utphys}	
\bibliography{paper}	

\end{document}